\documentclass[conference]{IEEEtran}

\usepackage{amsmath}
\usepackage{blindtext, graphicx}
\usepackage{algorithm}

\usepackage{bbm}

\usepackage{algpseudocode}
\usepackage{pifont}
\usepackage{tikz}
\usepackage{amssymb}
\usepackage{upgreek}

\usepackage{mathtools}

\begin{document}

\title{Multi-Agent Q-Learning for Minimizing Demand-Supply Power Deficit in Microgrids}

\author{\IEEEauthorblockN{Raghuram Bharadwaj D.}
	\IEEEauthorblockA{IISc, Bangalore, India\\
	 raghu.bharadwaj@csa.iisc.ernet.in}
	\and
	\IEEEauthorblockN{D. Sai Koti Reddy  }
	\IEEEauthorblockA{IISc, Bangalore, India \\
		 danda.reddy@csa.iisc.ernet.in}
	\and
  \centering	\IEEEauthorblockN{Shalabh Bhatnagar}
	\IEEEauthorblockA{IISc, Bangalore, India\\
		shalabh@csa.iisc.ernet.in}}

\maketitle

\begin{abstract}

We consider the problem of minimizing the difference in the demand and the supply of power using microgrids. We setup multiple microgrids, that provide electricity to a village. They have access to the batteries that can store renewable power and also the electrical lines from the main grid. During each time period, these microgrids need to take decision on the amount of renewable power to be used from the batteries as well as the amount of power needed from the main grid. We formulate this problem in the framework of Markov Decision Process (MDP), similar to the one discussed in \cite{goodmdp}. The power allotment to the village from the main grid is fixed and bounded, whereas the renewable energy generation is uncertain in nature. Therefore we adapt a distributed version of the popular Reinforcement learning technique, Multi-Agent Q-Learning to the problem. Finally, we also consider a variant of this problem where the cost of power production at the main site is taken into consideration. In this scenario the microgrids need to minimize the demand-supply deficit, while maintaining the desired average cost of the power production.

\end{abstract}

\section{Introduction}

Electricity is one of the most important components of the modern life. According to a recent survey, there are a total of 18,452 un-electrified villages in India \cite{elec}. Providing electricity to these villages is difficult for a number of reasons. The village may be situated very far away from the main grid and it would be difficult to establish a direct electrical line between the main grid and the village. Also, due to increasing global warming, we want to make less use of fossil fuels for the generation of power. Our objective in this work is to provide a solution to electrifying these villages.

The concept of smart grid \cite{farhangi2010path}, is aimed at improving traditional power grid operations.  It is a distributed energy network composed of intelligent nodes (or agents) that can either operate autonomously or communicate and share energy. The  power grid is facing wide variety of challenges due to incorporation of renewable and sustainable energy power generation sources. The aim of smart grid is to effectively deliver energy to consumers and maintain grid stability.

A microgrid is a distributed networked group consisting of renewable energy generation sources with the aim of providing energy to small areas. This scenario is being envisaged as an important alternative to the conventional scheme with large power stations transmitting energy over long distances. The microgrid technology is useful particularly in the Indian context where extending power supply from the main grids to remote villages is a challenge. While the main power stations are highly connected, microgrids with local power generation, storage and conversion capabilities, act locally or share power with a few neighboring microgrid nodes \cite{weiss1999multiagent}.
Integrating microgrids into smartgrid poses several technical challenges. These challenges need to be addressed in order to maintain the reliable and stable operation of electric grid.

Research on smartgrids can be classified into two areas -  Demand-side management and Supply-side management. Demand side management (DSM) (\cite{logenthiran2011multi, wang2010demand,dsm1,dsm2,dsm3,dsm4}) deals with techniques developed to efficiently use the power by bringing the customers into the play. The main idea is to reduce the consumption of power during peak time and shifting it during the other times. This is done by dynamically changing the price of power and sharing this information with the customers. Key techniques to address DSM problem in smart grid are peak clipping, valley filling,  load shifting \cite{maharjan2010demand}. In \cite{reddy2011learned}, Reinforcement Learning (RL) is used in smart grids for pricing mechanism so as to improve the profits of broker agents who procure energy from power generation sources and sell it to consumers.  

Supply-side management deals with developing techniques to efficiently make use of renewable and non-renewable energy at the supply side. In this paper, we consider one such problem of minimizing Demand-Supply deficit in microgrids. In \cite{goodmdp}, authors developed an MDP and applied Dynamic optimization methods to this problem. But when the model information (the renewable energy generation in this case) is not known, we cannot apply these techniques.

In our current work, we setup microgrids closer to the village. These microgrids has power connections from the main grid and also with the batteries that can store renewable energy. Owing to their cost, these batteries will have limited storage capacities. Each microgrid needs to take decision on amount of renewable energy that needs to be used at every time slot and the amount of power that needs to be drawn from the main grid. Consider a scenario where, microgrids will use the renewable energy as it is generated. That is, they do not store the energy. Then, during the peak demand, if the amount of renewable energy generated is low and power obtained from the main grid is also low, it leads to huge blackout. Thus, it is important to intelligently store and use the renewable energy. In this work, we apply Multi-agent Q-learning algorithm to solve this problem. 

\section*{Organization of the Paper}	
The rest of the paper is organized as follows. The next section describes the important problems associated with the microgrids and solution techniques to solve them. Section III presents the results of experiments of our algorithms. Section IV provides the concluding remarks and Section V discusses the future research directions.

\section{Our Work}
In this section, we discuss two important problems associated with the microgrids. We first formulate the problem in the framework of Markov Decision Process (MDP), similar to the one described in \cite{goodmdp}. We then apply cooperative Multi-Agent Q-Learning algorithm to solve the problems.

\subsection{Problem 1 - Minimizing Demand-Supply Deficit}

Consider a village that has not been electrified yet. We setup microgrids close to the village, and provides electrical connections to the village. It also has power connections from the main grid. Decision on power supply is generally taken in time slots, for example every two hours. Microgrids do not have non-renewable power generation capabilities. They have access to the renewable power and needs to obtain the excess power from the main grid. For ease of explanation, we consider two renewable sources - solar and wind. At the beginning of each time slot, microgrids obtain the power demand from customers with the help of the smart meters \cite{smart}. These microgrids need to take decisions on amount of renewable power to be used so that the expected long term demand-supply deficit is minimized.

One natural solution without the use of storage batteries is to fully use the solar and wind power generated in each time slot. The excess demand will be then requested from the main grid. If the requested power is less than or equal to the maximum allotted power, the main grid transfers this to the microgrids. Other wise, it transfers the maximum allotted power. This idea is described in Algorithm 1.

\begin{algorithm}
\caption{}
\label{greedy}
\begin{algorithmic}[1]
	\State Demand $\leftarrow$ Initial Demand
    \State get\_Demand(Demand) : Function that gives next Demand based on the current Demand.
    \State solar() : Function that computes solar power generated in a time slot
    \State wind() : Function that computes wind power generated in a time slot
    \State max\_prod : Maximum power alloted by the main grid.
    \State Deficit = 0 ; iterations = 0
    \Procedure{At every Time Slot}{}   
     \State $solar\_action = solar()$
     \State $wind\_action = wind()$
     \State $main\_action = max(0,min(max\_prod, Demand - (solar\_action+wind\_action)))$
     \State $Demand \leftarrow get\_Demand(Demand).$
     \State $Deficit += Demand - (solar\_action + wind\_action + main\_action)$
     \State $iterations += 1;$
    \EndProcedure
    \State $Average Deficit = Deficit/iterations;$
    
\end{algorithmic}
\end{algorithm}

Our objective in the current work is to do better than the above described solution. Microgrids are equipped with batteries that can store the renewable power. That is, we deploy the microgrids at the sites where there is availability of renewable resources. The goal is to intelligently use the stored power in batteries and take the optimal action at every time slot to minimize the demand-supply deficit.

We formulate the problem in the framework of Markov Decision Process (MDP). MDP \cite{sutton} is the most popular mathematical framework for modeling optimal sequential decision making problems under uncertainty. A Markov decision process is defined via tuple $< S; A; P;R >$, where S is the set of states. A is the set of actions. P respectively the probability transition matrix. $P_{a}(s,s^{'})$ is the probability that by taking action $a$ in the state $s$, the system moves into the state $s^{'}$. R is the reward function where $R_{a}(s,s^{'})$ is the reward obtained by taking action $a$ in the state $s$ and the $s^{'}$ being the next state. In infinite horizon discounted setting \cite{vol1}, the goal is to obtain a stationary policy $\pi$, which is a mapping from state space to action space that maximizes the following objective:

\begin{equation}
\sum_{t = 0}^{\infty} \gamma^{t}R_{a_{t}}(s_{t},s_{t+1}),
\end{equation}
where $a_{t} = \pi(s_{t}),$ and $\gamma \in [0,1]$ is the discount factor.

We model Power Demand as a Markov process. That is, the current Demand depends only on the previous value. The Demand value, and the amount of power left in the batteries form the state of the MDP. That is, 
\begin{equation}
 state = [Demand,\hspace{0.1cm} solar\_batterylevel,\hspace{0.1cm} wind\_batterylevel].
\end{equation}

Based on the current state, the microgrids need to take decision on the number of units of power to be used from their batteries, and the number of units of power that is needed from the main grid. Note that the power from the main grid is fixed and bounded during all the time periods, while power from the  renewable sources is uncertain in nature. 

\begin{equation}
Action = [solar\_power,\hspace{0.1cm} wind\_power,\hspace{0.1cm} main\_power],
\end{equation}

The state evolves as follows:
\begin{equation}
\begin{split}
solar\_batterylevel &= solar\_batterylevel - solar\_power \\&\hspace{1cm}+ solar(), 
\end{split}
\end{equation}
and
\begin{equation}
\begin{split}
wind\_batterylevel &= wind\_batterylevel - wind\_power 
\\&\hspace{1cm}  + wind(),
\end{split}
\end{equation}
where
$solar()$ and $wind()$ are the solar and wind power generation policies. 
Our objective is to minimize the expected long term discounted difference in the demand and supply of the power. So, the single stage Reward function is as follows:

\begin{equation}
\begin{split}
R = -&(Demand - \\
&(solar\_power+wind\_power+main\_power))^2
\end{split}
\end{equation}
When there is uncertainty in the system, like the renewal energy generation in this case, traditional solution techniques like value iteration and policy iteration cannot be applied. Reinforcement Learning provides us with the algorithms which can be applied when the model information is not completely known. One such popular model-free algorithm is Q-learning \cite{vol2}. However, microgrids applying Q-Learning independently is not a feasible solution, as they are working towards a common goal. Hence we apply co-operative Multi-Agent Q-Learning algorithm \cite{marl} to solve the problem. At every time slot, the microgrids exchange the battery level information among themselves. Then they apply Q-Learning on the joint state and obtain the joint action. Each microgrid then selects its respective action. Let the joint state at time $k$ be $s_{k}$. We select a joint action, based on an $\epsilon-$greedy policy. An $\epsilon-$greedy policy is one where we select the action that gives maximum reward with probability $1-\epsilon$ and a random action with probability $\epsilon$. Let it be $a_{k}$ and so we move to a new state $s_{k+1}$. Let the single stage reward obtained be $r_{k}$ Then the Q-values are updated as follows:
\begin{equation}\label{Q-val}
\begin{split}
Q(s_{k},a_{k}) = Q(s_{k},a_{k}) + \\
 \alpha[r_{k}+\gamma max_{a}Q(s_{k+1},a) - Q(s_{k},a_{k})], 
\end{split}
\end{equation}
where $\alpha \in [0,1]$ is the learning parameter and $\gamma \in [0,1]$ is the discount factor.

We update this Q-values until convergence. At the end of this process, we will have an optimal policy that gives for each state, the optimal action to be taken. The idea is described in  Algorithm 2. 

\begin{algorithm}
\caption{}
\label{q-learning}
\begin{algorithmic}[1]
	\State  State $\leftarrow$ Initial State 
    \State get\_Demand(Demand) : Function that gives next Demand based on the current Demand.
   \State solar() : Function that computes solar power generated in a time slot
    \State wind() : Function that computes wind power generated in a time slot.
    \State max\_prod : Maximum power alloted by the main grid.
    \State Deficit = 0 ; iterations = 0
    \Procedure{At every Time Slot and for all the microgrids}{}   
     
     \State Select an joint action $a$ vector according to $\epsilon$-greedy policy.
     \State $new\_solar = solar - a(solar\_action) + solar()$
     \State $new\_wind = wind - a(wind\_action) + wind()$
     \State $new\_Demand = get\_Demand(Demand)$
     \State Update the Q-values according to \eqref{Q-val}
     \State $state = [new\_Demand, new\_solar, new\_wind]$
     \State $Deficit += Demand - (a(solar\_action) + a(wind\_action) + a(main\_action))$
     \State $iterations += 1;$
    \EndProcedure
    \State $Average Deficit = Deficit/iterations;$
\end{algorithmic}
\end{algorithm}

\subsection{Problem 2 - Balancing Demand-Supply Deficit and cost of Power Production}
In the above formulation, cost of the non-renewable power production at the main grid site is not taken into consideration. Therefore, it is natural to use the maximum alloted power from the main grid and the optimization is done on the amount of power drawn from the solar and the wind batteries. In this formulation, we put a cost on amount of power that can be obtained from the main grid. That is, we modify our single stage Reward as follows:
\begin{equation}\label{prob2}
\begin{split}
 -c*(Demand - (solar\_power + wind\_power+ \\
 main\_power))^2+(1-c)*(main\_power)^2.\\
 c \in[0,1],
 \end{split}
\end{equation}

where $c$ is the parameter that controls the demand-supply objective and the main grid production cost. Note here that when $c =1$, it is same as the problem 1. In this formulation, our objective is not only to minimize the demand-supply deficit, but also maintain a desired average production of power at the main site. Similar to the above problem, we apply multi-agent Q-Learning to obtain the optimal solution. 

\section{Experiments}
We consider two microgrids operating on the solar and the wind power batteries respectively. Demand values are taken to be 8,10 and 12. The probability transition matrix for the Demand is given below. 

\begin{align}
 P =  \left [\begin{array}
{ccc}0.1 & 0.6 & 0.3 \\
     0.3 & 0.1 & 0.6 \\
     0.6 & 0.3 & 0.1 \\
\end{array} \right].
\end{align}

The maximum storage capacity of the batteries is set to 5. The renewable energy generation process for simulation purposes is taken to be Poisson with mean 2. Discount factor of the Q-learning algorithm is set to 0.9.

We begin our simulations with the state $[8,5,5]$. First we run Algorithm 1 for $10^6$ iterations and compute the average deficit in demand and supply. Then we run Algorithm 2 for $10^8$ iterations with $\epsilon-$greedy policy with $\epsilon = 0.85$ and obtain an optimal policy. We then compute the average deficit using optimal Q-values for $10^4$ iterations. We compare both the algorithms in Figure 1.

\begin{figure}[h!]
\begin{center}
 \includegraphics[width = 10cm, height = 8cm]{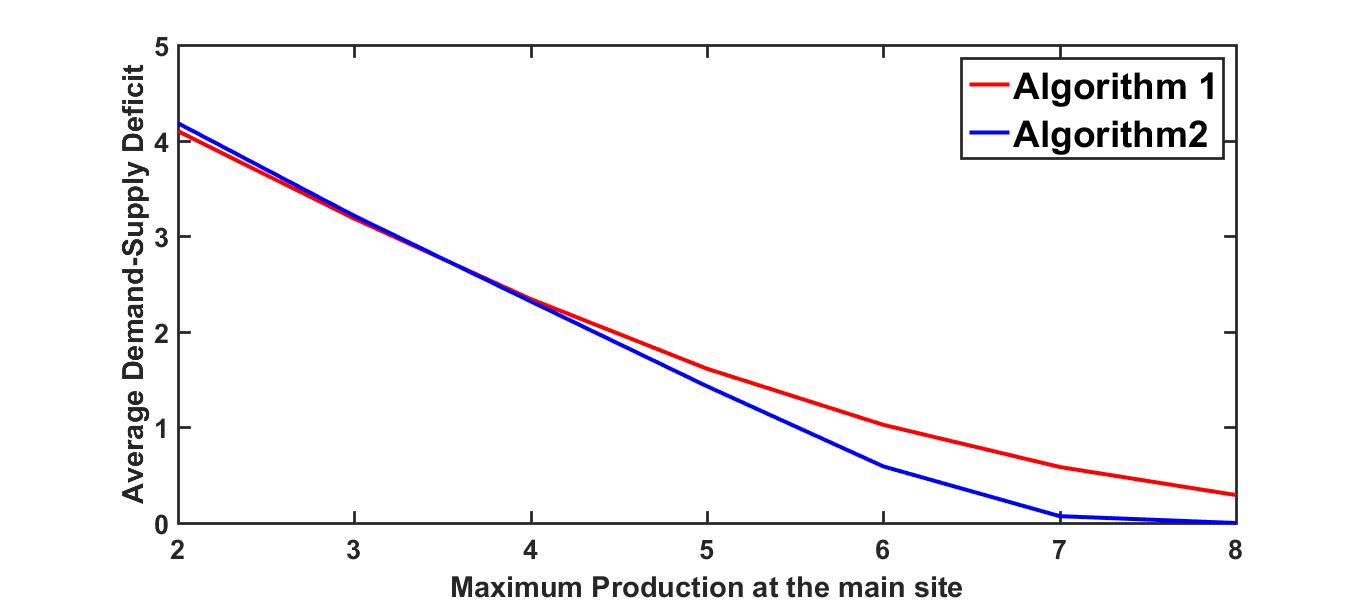}
 \caption{Comparison of Algorithms on Problem 1}
 \end{center}
 \end{figure}

With regard to the problem 2, we plot the values of average deficit in the power and the average production at the main grid obtained by different values of $c$ in \eqref{prob2}. The maximum power allocation at the main site is taken to be 8. This is shown in Figure 2. 

\begin{figure}[h!]
\begin{center}
 \includegraphics[width = 10cm, height = 8cm]{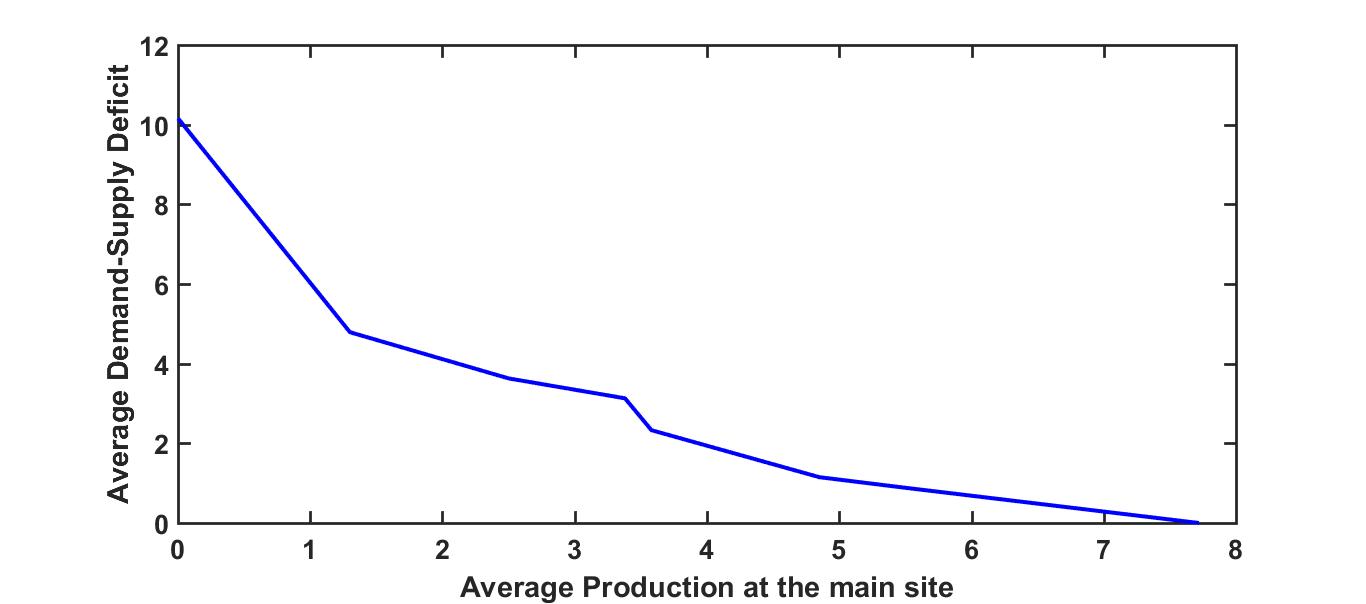}
 \caption{Algorithm 2 applied to Problem 2}
 \end{center}
 \end{figure}

\subsection{Observations}

\begin{itemize}
\item In Figure 1, we see that the Algorithm 1 performs better than the Algorithm 2, when the Maximum Power Allocation (MPA) at the main site is less than 3. This is because, when the MPA is very less compared to the minimum demand value, storing the power in the batteries doesn't provide any advantage. 
\item On the other hand, if the MPA is not very small and comparable to the minimum demand, the Algorithm 2 outperforms Algorithm 1. Thus, in this case it is useful to store the power in the batteries. This is the main result of our paper. 
\item In Figure 2, we observe that as the value of $c$ increases, the average demand-supply deficit decreases and the average power obtained from the main grid increases. 
\end{itemize}

\section{Conclusion}
We considered the problem of electrifying a village by setting up microgrids close to the village. These microgrids have access to the renewable energy storage batteries and also electrical connections from the main grid. We identified two problems associated with the microgrid. First problem is to minimize the expected long-run discounted demand-supply deficit. We model this problem in the framework of MDP \cite{goodmdp}. This formulation doesn't take into consideration the cost of power production at the main grid site. Finally, we formulated MDP taking the cost of power production into consideration. We applied Multi-Agent Q-Learning algorithm to solve these problems. Simulations show that, when maximum power allocation at the main site is not very less, storing the power in the batteries and using them intelligently is the better solution compared to not using the storage batteries. 

\section{Future Work}
As future work, we would like to consider the possibility of power sharing between the microgrids. In this case, along with decision on amount of power to be used from stored batteries, microgrids also have to make decision on the amount of power that can be shared with others. We would also like to consider the  heterogeneous power price system. In this scenario, the price of power production at the main grid will vary from time to time. 

\section*{Acknowledgment}

The authors would like to thank Robert Bosch Centre for Cyber-Physical Systems, IISc, Bangalore, India for supporting part of this work.

 \bibliographystyle{IEEEtran}
 \bibliography{IEEEabrv,reference}

\end{document}